# Social Networks and Collective Intelligence: A Return to the Agora

**Manuel Mazzara**
*UNU-IIST, Macau and Newcastle University, UK*
**Luca Biselli**
*Independent Researcher, UK*
**Pier Paolo Greco**
*Newcastle University, UK*
**Nicola Dragoni**
*Technical University of Denmark, Denmark*
**Antonio Marraffa**
*Polidoxa.com, Germany*
**Nafees Qamar**
*UNU-IIST, Macau*
**Simona de Nicola**
*University of Bologna, Italy*

**ABSTRACT**

Nowadays, acquisition of trustable information is increasingly important in both professional and private contexts. However, establishing what information is trustable and what is not, is a very challenging task. For example, how can information quality be reliably assessed? How can sources credibility be fairly assessed? How can gatekeeping processes be found trustworthy when filtering out news and deciding ranking and priorities of traditional media? We are studying an Internet-based solution to a human-based ancient issue and we call this solution Polidoxa, from Greek "poly" (πολύ), meaning "many" or "several" and "doxa" (δόξα), meaning "common belief" or "popular opinion". This old problem will be solved by means of ancient philosophies and processes with truly modern tools and technologies. This is why this work required a collaborative and interdisciplinary joint effort from researchers with very different backgrounds and institutes with significantly different agendas. Polidoxa aims at offering: 1) a trust-based search engine algorithm, which exploits stigmergic behaviours of users' network, 2) a trust-based social network, where the notion of trust derives from network activity and 3) a holonic system for bottom-up self-protection and social privacy. By presenting the Polidoxa solution, this work also describes the current state of traditional media as well as newer ones, providing an accurate analysis of major search engines such as Google and social network (e.g., Facebook). The advantages that Polidoxa offers, compared to these, are also clearly detailed and motivated. Finally, a Twitter application (Polidoxa@twitter) which enables experimentation of basic Polidoxa principles is presented.

## INTRODUCTION

In the democratic city state of ancient Greece, the "agora" (from Greek: Ἀγορά, "gathering place" or "assembly") was the place where citizens used to meet, discuss, exchange information and make important decisions about the future of society. In this place, the concept of public sphere was born: in fact it was considered another kind of space, a sort of empty space next to the private one.

The citizens used to meet there and considered it not a personal but a common space. It was by virtue of this effort that the perfect model of a democratic city was born: the agora was the meeting place of the assembly of citizens: in the public square each person is equal no one is subjected to any other. The agora was the political place of a multitude, composed by different parts but similar at the same time, all with the same rights. It was a place where discussions occurred without violence, force and abuses. Hannah Arendt in Vita Activa (Vita Activa, 1958) identified in this Greek model of cities the highest forms of citizenship: every Athenian citizen, in person and not through representation, when a serious risk and danger occurred, used to go in the agora and discuss about the highest issues, committing themselves to put into practice what has been said. This was a political system based on equality of knowledge, information exchange and decision making fairness. Nowadays, the mechanism by which information is spread across (and consequently how decisions are made) has had a significant change in nature. In fact, the majority of people retrieve their information from major TV stations, radio and newspapers. The weakness of this mechanism is that it is a one-way information, not a cross-flow one. This means that citizens have lost their ability to interact with the decision making process. Consequently, the concept of "agora" is lost in favour of a different mechanism.

These days the average citizen gets access to information mainly by watching TV, especially the main national channels. Radio, newspapers and magazines represent a secondary source of information but they are hardly comparable to the power of TV. In particular, reading takes time and it does not suit well the hectic life style of modern times. As a consequence, information obtained by reading books can be considered quite negligible for an adult citizen with an average level of education. Another major problem comes from the fact that the majority of the world population speaks just its native language while some information is not always accessible in that language. Furthermore, to have a complete unbiased (or at least, multi-biased) source of information, it would be quite useful to access documents coming from diversified sources in different languages. According to the A.C. Nielsen Co, the average American citizen watches more than 4 hours of TV each day (or 28 hours/week, or 2 months of nonstop TV-watching per year). In a 65-year life span, that person will have spent 9 years watching TV. The percentage of Americans regularly watching TV while eating dinner is 66%, while 49% say they simply watch too much of it (Television Statistic, in "The Source book for teaching science", www.csun.edu/science/health). These are very alarming numbers and they may raise health concerns, but the authors believe that there is an even bigger issue behind them. Accessing information mainly or exclusively from TV, as the common experience (plus statistics) shows, is obscuring the potential of other sources of information like radio, newspapers, magazines, books, the Internet or a community of trusted contacts. These other sources are generally able to provide a much wider opinion range. Indeed, we are not really able to access unbiased sources, but we could get what we call a "multi-biased source" at least. A more heterogeneous set of different viewpoints, which then could stimulate human critical thinking and cognitive interpolation is desirable.

## Gatekeeping Process

The fundamental problem with TV news is that the information streaming is simply unidirectional, i.e., there is no possibility for the audience to interact with the process in any way. This is clearly the opposite principle with respect to that of the "agora". The final result obtained from a mass media passes through many levels of organizational processing on its way to the audience and, at each step the original data is filtered – reduced in length, edited and so on. Each step in the process could be identified as a gate, through which the data must pass before reaching the final users. Consequently, this situation is known as gatekeeping (White, 1950, Mccombs & Shaw, 1972. See Figure 1). Gatekeeping is generally a very effective mechanism to ensure that an irrelevant or misleading piece of information will not reach the general public. It determines a *quality assurance* process and an expert evaluation similar to what happens in conferences/journals peer-reviewed system. However, there is also a potential drawback: with TV and its gatekeeping mechanism, the audience is not able to provide a real time feedback, and this may cause misunderstandings and lack of active interaction. Furthermore, people are not able to decide their information source or the type of content or express the will to expand some topics either. This means that mass media tend to set the "agenda", i.e., the list of items that people will be discussing. This theory is known as agenda-setting theory (McCombs, Shaw, 1972 & McCombs, 2004) and asserts that mass media have a large influence on audiences, choosing which stories have to be considered newsworthy and how much prominence and space they are given. Agenda-setting theory's main postulate is salience transfer *(ibidem)*. Salience transfer is the ability of a mass media to transfer relevant issues from their news media agendas to public agendas. Thus, the power of the media may lie not in its ability to determine people's opinions, but rather in its role of determining what issues will be considered important enough to discuss. Whatever is not appearing on the main media simply does not exist. This has a quite subtle consequence. The German political scientist Elisabeth Noelle-Neumann has defined an important theory called the spiral of silence theory (Noelle-Neumann, 1974). This theory asserts that a person is less likely to voice an opinion on a topic if he/she feels that idea belongs to a minority. This is for fear of reprisal or isolation from the majority. Thus, TV news can easily transfer this feeling to the watcher who is following the news from his home, maybe at a time of the day when the attention is not at its peak. As stated earlier, 66% of Americans regularly watch television while eating dinner and this is the time when news are usually broadcasted (McCombs, Shaw, 1972). A relevant social experiment emphasizing the fragility of the person in a mass society when he is confronted with the contrary opinion of a majority, and the tendency to conform even if this means to go against the person's basic perceptions, was made by Asch (Asch, 1955). Entman also described the spectator's condition very well in his article "How the media affect what the people think" where he explains how media contribute to what people think precisely by affecting what they can think about (Entman, 1989).

*Figure 1. Gatekeeping.*

In some cases, the fact that information goes through gatekeeping (i.e., every journalist has to go through several levels of approval like director, editor, company shareholders before the information is released to the public) could lead to situations that are unfavourable to the final "consumer". Consider, for example, the case in which news agencies are purchased and they become part of a larger business, where providing information may not be the main core business or in addition they could be affected by the company's position on the Stock Market. Example of this has been the concern that Reuters' objective reporting may be affected by recent merging with Thomson Corporation, owning 53% of the company, in contrast with the 15% limitation to share ownership historically imposed by its constitution to preserve freedom and integrity of the agency. More information on this concern can be found on the bbc website (http://news.bbc.co.uk/2/hi/business/6656525.stm).

Once the gatekeeping process has been understood, its potential risks and limitations have to be accepted together with its advantages. Now, if we consider how the main channels and news agencies are more and more centralized (like every other business), it is not difficult to realize how the whole mass media communication sector has the potential to be put under control in the future, especially in some countries where the democratic process is considered weaker (Maurer & Kolbitsch, 2006).

**Outline and Contributions**
This work contributes with several principles and technicalities to build a social platform to achieve collective intelligence via information sharing among trusted contacts. It also presents a Twitter-based implementation of a subset of these principles. Trust modeling, social networks, collective intelligence, algorithms and the relative motivations supported by literature in communication sciences are a quite inedited interdisciplinary blend, which has not been really investigated so far. We intend to pursue our investigation and expand our knowledge on this topic.

The chapter is structured as follows: after an introduction on the problems and motivations which led to this research, an overview on traditional media, search engines and social networks is presented and a synopsis is offered. What Polidoxa is and how its trust model works is then detailed. Specifically, the concept of trust is investigated under different perspectives and as a function of several parameters. Privacy and security, among others, are also considered. The relevant concept of "immunity" is also investigated. Finally, a Twitter-based implementation is presented as a first proof-of-concept prototype of the Polidoxa platform.

**TRADITIONAL MEDIA**
The mechanism by which traditional media (like TV, radio and newspapers) intrinsically operate, is to allow only passive actions, i.e., reading, watching and listening to specific contents according to the opinion of some expert or authority, which should guarantee the quality of information. The audience here does not control the medium content, the agenda setting, and the choice of experts and commentators in charge of presenting the facts. This means that the media owner (the publisher) indirectly chooses who the experts are and controls who says what. Indeed, in a globalized world, media from different cultural, political or religious background, present quite different interpretations of facts coming from different "experts". With so many sources of information and no shared and agreed evaluation parameters to decide who is an expert and on what, citizens are left in confusion. Even when, for fairness (or in Latin "par condicio"), experts from different parts are involved in the discussion, the user has still no chance to intervene in the process. The only freedom and choice given to the audience is switching the media, or that specific channel, off. The Communication model is consequently unidirectional and it relies on three rigid rules: Gatekeeper, Speaker and audience, as shown in Figure 1.

## INTERNET AND SEARCH ENGINES

Given the limitations of the traditional media, it is necessary a look into new media to identify how the gap could be filled. Since Internet offers an open platform to exchange information and can be considered a paradigm shift similar to the one that was brought by the Gutenberg's invention of mechanized press, it seems to be a valid target for this research.

With the arrival of Internet, the limitations of traditional media (i.e., offering only the passive actions of reading, watching and listening) can be overcome. It is indeed possible for users to control the information they achieved, choose the content to read and how to interact with other users or bloggers. It is also possible to choose the timing for accessing the information and by doing so, the tendency of watching the news while dining, which coincides with the weakest time for a critical thinking process, can be avoided. Internet has the full potential to reverse (or at least minimize) the process, avoiding the agenda-setting theory issues or the spiral of silence condition. However, to exploit this potential, users need some know-how: given the limited control on the information on the Web, it is possible to find very good pieces of so-called "alternative journalisms" as well as any kind of hoax.

Internet is not a passive media like TV: users are expected to be active and critical thinking is stimulated. However, users have to be educated to use the media. The potential of Internet could be seriously reduced in the future, if focal nodes will be set under control with the same gatekeeping process discussed for the traditional media. Again, gatekeeping is effective at ensuring quality but it limits feedback process and critical thinking. To the best of our knowledge, search engines like Google or social networks like Facebook are, for most users, the starting point of the majority of their research. So the important question is: how can we be sure that these nodes are trustworthy? Let us briefly analyse the main characteristics of these two powerful instruments in the hands of Internet users.

### Search Engines

Today the most popular and used search engines on the market allow users to search over nothing less than trillions of different documents. Such a possibility was totally unthinkable only 20 years ago. However, results coming from these engines are mainly commerce-oriented and purely based on a quantitative algorithm that has significant margins of improvements in terms of results quality. For example, simply typing in Google bar "economic crisis in Europe" we are offered millions of links and their order is purely decided on the basis of the most linked pages, which tells nothing about the specific user needs, which are different from one person to another. There are two critical aspects of Google ranking: first, pages coming from popular newspapers are top ranked apart from their content; second, privileged ads slots can be bought, again independently from their content. Consequently, gatekeeping process is still an open issue and Google could be influenced as easily as TV channels. Furthermore, the communication model is still unidirectional. Given the enormous amount of information available on the Web and the typical user profile and effort put into the search, vast amount of information remains, in fact, inaccessible for users. Therefore, this communication model remains not very different from the one offered by traditional media, i.e., the Gatekeeper (databases and ranking algorithm), Speaker (result page), Audience (users). The major lack in this model is the fact that the audience has no chance to self-configure the ranking algorithm and therefore being able to influence the content and the order of the search results (apart from minor flexibility given by the use of advanced search features). The way in which the search engine presents the results still remains a black box for the average user. Overall, it could be said that "what is not found on the first page of

Google results does not exist". Considering how powerful a medium like Internet is, we would expect users to be somehow able to interact with it in a much more proactive way.

## SOCIAL NETWORKS

Social networks service like Facebook have a focus on collecting and sharing users data (family, friends, pictures etc...) and personal interests/information. These days, they have massive numbers of users accounting worldwide for an incredible amount of hours of usage. If we disregard all the private information posted on Facebook, which are irrelevant for people outside your closest group, the platform can be proactively used to share quality information. Indeed, social networks are very different from search engines, because of the way the source of information can be controlled by users. A generic user, for example, can follow a specific trusted friend or hide information coming from (or going to) untrusted users, who seem to post information considered irrelevant. Unfortunately, even with Facebook, users are not able to rank information since all posts are only shown chronologically. Users are not able to set content alerts to be informed only about specific topics. Another issue is that users cannot enrich their posts by linking pieces of information that are not on the Internet, although this is becoming less and less relevant as all the other media are also posting their contents on the Web.

Being Facebook supported by advertising, this information is more critical in term of quantity, rather than quality. The way Facebook is structured does not consequently promote or improve critical thinking, learning, comprehension and discussion among its users. Mechanisms such as "like", for example, are structured for giving just a quick evaluation, which, as a consequence, may be simply an accelerated feeling, not moderated by critical thinking. According to a Nielsen's Company research, people are spending more and more time on social networks. Global average time spent is in fact about five and half hours per month and this number is increasing, with Facebook currently dominating its position as a destination (http://mashable.com/2010/03/19/global-social-media-usage/).

Social networking is globally expanding and it is likely to deeply influence the way people will interact with other people in the future, promoting connections able to go beyond the classical geographical limits (Mazzara, Marraffa, Biselli, Chiarabini, 2011). At the same time, social networks have some other interlinked privacy and security issues that are discussed hereunder.

## TRUST, PRIVACY AND SECURITY IN SOCIAL NETWORKING

Massive use-based growth of social networks such as Facebook, LinkedIn, and MySpace --with millions of users everyday-- play an indispensable role in our daily-life. Nonetheless, dissemination of unverified (dis)information on such social networking sites can affect individuals as well as religious, ethnic and geographically dislocated communities. Such a factor alludes to intentional- or unintentional sharing of information where people are pulled into fierce debates as well as conflicts. This is actually a "wild expansion" of social networks (in terms of data) that has paved the way to share personal opinions along with possibly forged facts favouring a specific group. Albeit social networks proclaim several benefits and competitive features, they are not exempt from subtle data leakages and facts misinterpretations. This further means that there is no independent source to confirm the validity of given facts and figures. They also lack in specifying and implementing appropriate security and privacy procedures to protect users' data. For example, our whole Facebook album is exposed to a stranger after a comment from a friend in our network. Also, how our stored information can be manipulated is one of the unknown aspects. Several famous personalities across the globe have been the victims of intruders and attackers on such social networks. Ordinary users are generally unaware of such intricacies unless they encounter certain inconveniences against which they report loss of data and misuse of their accounts. Such victims are large in numbers and they need coordinated efforts to deal with their issues.

Yet another and implicit aspect of users' data may as well be sold by a social network to a third party. An outcome can also be compromising data integrity and confidentiality (irrespective of the intentional or unintentional move) due to flexibly implemented security mechanisms or inadequate security policies. These issues are encouraging to underpin the state-of-the-art trust, security and privacy, and the collaborative ability among users of such social networks. The alluded issues have been already approached in various ways such as Safebook (Cutillo, Molva, Önen, 2011), which offers a so-called replacement to Facebook using P2P network in a more decentralised way. Safebook puts a special emphasis on the privacy of its users with regards to the application provider and shields against malicious users or intruders. Ding, Cruz and Li (2009) attempted to model a feature social network called friend suggestion. Their approach is based on high level Petri nets, but extended with channels to formally model social networks. Another recent added challenge to security and privacy is in mobile social networks, which require user's location and preferences. Issues reported in mobile social networks can be found in (Beach, Gartrell, Han, 2009) such as direct anonymity issues and eavesdropping, spoofing, replay and wormhole attacks. This lack of security and privacy does not surprise at all, since social network applications do not take into account security and privacy by design.

To the authors' interest, trust and a cohesive collaboration environment are vital to understand and enhance social networks' security and collaboration on the available information. To this end, use of formal methods in terms of analysing and reasoning security and privacy properties based on the proposed trust model are justified as they allow simulating patterns and systems existing in nature. For example, using formal methods one can unambiguously specify pattern of social relationships and then reason about it. For instance, (Fong, Anwar & Zhao, 2009) have provided a formalized model of Facebook access control mechanism and reported it as a Discretionary Access Control (DAC).

Our proposed construction of formal models for modeling trust in social networks requires a rigorous treatment by using novel concepts and then by allowing formal reasoning over the constructed models. However, before constructing trust model one should think of access control mechanism to be applied. For example, access control mechanism and a security policy play the role of backbone in such systems which can be taken into account to construct formal models. For instance, DAC can only be used where the user of a computer system is fully aware of the consequences of a granted permission and revoking it, which is just not the case in Facebook. It is pertinent to note that although Facebook offers DAC it eventually fails to handover all the control to its users. Thus, a greater portion of the information is beyond the control of users, and actually the information is centrally administered -- just as in Role Based Access Control (RBAC). Centrally administered security policies result in exposing users' photo albums or a wall post, and similarly falsely suggesting a friend outside one's network.

The notion of trust is equally applicable to social networks. For example, Bonneau, Anderson and Church (2009) suggested having privacy suites that can be chosen from the user's privacy settings. These suites would be specified by friends or trusted experts, with the possibility to be modified by its user if necessary. One of the reasons to opt for such an option roots back to the problem that users lack in understanding privacy settings. For example, Facebook presents 61 settings on 7 different configuration pages, and LinkedIn has 52 settings on 18 Pages (ibidem). Thus, given a trust model between a user's friends and other experts, one can reduce the security and privacy threats in social networks. However, those are unformalized trust models and incomparable to our approach in that they do not study and embed the mechanism in addition to applying access control. Our extended focus is also to monitor the information flow on these social networking sites and to address users' needs accordingly from media perspectives.

## SYNOPSIS
In synthesis, our research identified the following:

- Traditional media: the content is controlled by the gatekeeper.
- Web and search engines: the content is controlled by the gatekeeper, but users can decide the topic. However, the requested content has to be stored in the corporation databases and this content has to appear reasonably high in the engine ranking to be accessible to the average user.
- Social networks: the content is not controlled by any central authority or gatekeeper, but it is controlled by the specific user belonging to a contacts' network. The major feature of a user's network is trustworthiness of the content.

## POLIDOXA AND TRUST

The advent of social networks may give rise to a paradigm shift in communication provided that a number of issues are solved. Our objective are those of combining the potential of search engines to quickly retrieve information and the ability of controlling its source, which is typical of social platforms. Polidoxa (from Greek "poly", (πολύ), meaning many or several and "doxa" (δόξα), meaning "common belief" or "popular opinion") is a platform which aims at introducing the concept of "trust" in social networks to improve information quality and general knowledge. In social sciences "trust" is defined as a situation where one party is willing to rely on the actions of another party (Mayer, Davis, Schoorman, 1995).

More formally, let us define a set U of users; the function *trust* is defined as follows:

*Trust: (A ϵ U, B ϵ U)* → *[0,99]*

That means, the trust of a user A for user B is expressed by a natural number between 0 and 99. For example, *Trust (Alice, Bob) = 99* means that Alice consider Bob a very trustable individual. It is worth noting that this function is not transitive, i.e., it can be that *Trust (Bob, Alice) = 0*.

At the moment, social networks like Facebook or LinkedIn allow only information to be shown chronologically or being filtered in some very basic way. There is no notion/acknowledgement of "trust" between users and different contacts have similar relevance. Polidoxa is instead based on the principle that immediate contacts have more influence, while the others see a reduction of their influence which is proportional to their distance. Even direct contacts are not all at the same level, but users can decide a "trust" score and this score will change over time according to their activities. Polidoxa is based on the principle of collective/swarm intelligence which is the normal way of operating between colonies of insects living in collaborative communities (Maurer & Kolbitsch, 2006; Joslyn, Rocha, Smith, Johnson, Rasmussen & Kantor, 1998).Trust is the key to information and Internet has an enormous potential to fix the issue of information trustworthiness.

### Multi-Dimensional Trust

In the previous section trust was considered to be a mono-dimensional entity. In reality, trust between individuals is not a mono-dimensional entity, but a multi-dimensional one. Multi-dimensional trust can be formally defined as follows (U is a set of users and T is a set of topics):

*MTrust: (A ϵ U, B ϵ U, t ϵ T)* → *[0,99]*

That means, the trust of a user A for user B regarding a topic t is again expressed by a natural number between 0 and 99. For example, *Trust (Alice, Bob, football) = 99* means that Alice consider Bob a very trustable individual. It is worth noting that this function is not transitive, i.e., it can be that *Trust (Bob, Alice, football) = 0* while, at the same time, it can be *Trust (Bob, Alice, fashion): 99*.

A given topic *t* directly defines *a projection* of trust over a user's contacts.

*Experts: (A ϵ U, t ϵ T)* → *P(U)*.

For example, for Alice Bob, Ken and John are football experts and their opinion is highly valuable:

*Experts (Alice, football) = {Bob, Ken, John}*

Once a subjective set of experts for a given user and topic has been individuated, a number of analyses can be performed on these experts, for example opinion mining.

### Opinion Mining and Collective Intelligence

Although other researchers have used swarm intelligence techniques to get high quality data from web communities, applying swarm intelligence algorithms to social networks to achieve collective intelligence is an open research domain. One of the most promising investigations is described in "Swarm Intelligence for Analysing Opinions in Online Communities" by the University of Erlangen-Nuremberg in Germany. In this work, text mining techniques are combined to ant colony metaheuristic algorithm to perform opinion mining. This research can be divided in two major parts: 1) opinion mining and 2) use of ant colony for swarm opinion forecast. The main goal of this work is to distinguish between "positive", "negative" and "no opinion". The method consists in separating the words in each sentence and calculating the relative frequencies. At that point, polarity of each post is calculated. The results of this work are presented in Figure 2.

*Figure 2. Opinion mining.*

Once opinion mining has been performed, an algorithm inspired by the ant colony metaheuristic can be used. The actual implementation of the algorithm consists in using posts polarity as ant pheromones. In this way, ants can predict next post polarity. More details about ant algorithm will be given in the following sections.

## Quarantine and Trust as a Function of Distance

Trust is not only a multi-dimensional, but also a multi-level concept. Google+ (https://plus.google.com/), for example, evaluates only the first degree of separation between contacts. Polidoxa, instead, aims at evaluating the whole network of contacts, assuming knowledge sharing as being important even when coming from indirect sources. The assumption is that immediate contacts have more influence, while other contacts from different levels see a reduction of their influence, which is somehow proportional to their distance. Every user of Polidoxa has an inner circle of first contact users which he/she likes to follow and are considered information sources and generators. Whoever is not in this immediate set of trusted sources belongs to "the rest of the world", a grey mass of users about whom he/she does not have any information. Polidoxa aims at offering a second list of users, i.e., a "selection" of people from "the rest of the world" which has the potential to become relevant and trustworthy by the user. This list of people will be kept initially in a "quarantined mode", i.e., under observation and the user will be able to pick up some (or all) of those and bring them into the set of direct contacts. How do these candidates are selected from the system among the (potentially) millions of users? It is well-known that every person in the world is separated on average by anybody else by six steps; at least in western urban world. This fact is well known as the "Six degrees of separation theory" or "Small-world experiment" (Travers & Milgram, 1969). Thus, how can the "most trustable" persons in the system be suggested to the users? A mathematical model of "trust transitivity" needs to be developed. How does trust decrease when we pass from one level of separation to the next one? This issue is not entirely solved at the moment and several possible solutions are under consideration. The most obvious, simple, but imprecise solution is defining the inferred trust as decreasing in a linear way.

Let us define a function expressing the distance between two users:

*Dist: (A $\epsilon$ U, B $\epsilon$ U) = $\rightarrow$ N*

Now, let us suppose we have (with k=99 in this case):

*Dist(A,B) =x*

then

*Trust (A, B) = k for x=1*

*Trust (A, B) = 1/x * k for x>1*

This means that first level contacts have here a value trust of 99 and the indirect contacts see their trust decreasing in a linear way. Of course, this is a simplification since the direct contact trust can be in fact set by the user (in practice this k is changing over time, see the following sections to understand how further parameters are implied in this change).

However, this solution is imprecise because we know by experience that trust is not a linear relationship, i.e., the contacts a person has at the third or even fourth level, have a value which is generally close to zero while direct contacts or contacts of contacts are very valuable. Other better possibilities are expressed in Figure 3.

We are evaluating another ranking system based on a trust relationship inspired to a Kepler-Newton modeling system. During our life time we in fact trust our parents, relatives, friends, or even people we do not know, creating our solar system. We add "new planets" which we critically found compatible to the beliefs of our mental galaxy. Our contact links are based on a non-linear relationship, where the quality of trust increases when it gets closer to our beliefs, knowledge, commitment etc. Research in this area has been already developed at McGill University, Canada (Maheswaran, Tang, and Ghunaim, 2007). The Inverse Square Law on which the idea is based is shown in Figure 4. We can make a simpler analogy between this idea and how forces distribute over a sphere. By defining the intensity $i$ of the Trust as: $i = T/A$ where $T$ is Trust and $A$ the area of the sphere, i.e., our social network, we get $i = T/A = k*T/(4\pi x^2)$ with x the radius. Thus, if $x2 > x1$ then $i2 < i1$ which means the more the contact is distant, the less powerful the trust is.

*Figure . Trust definition.*

*Figure 4. Inverse square law.*

## SOCIAL NETWORKS, SWARMS AND COLLECTIVE INTELLIGENCE

A platform like Polidoxa, based on a community of a potentially significant number of members (the humankind) has the potential of becoming a "Swarm", using an analogy inspired by the concept of "swarm intelligence" found in Nature (Beni, Wang, 1989). We have studied, at a micro scale level, whether "Swarm intelligence" can help parts of Polidoxa algorithms and, at macro scale level, whether new interesting aspects can emerge regarding intelligence and knowledge sharing coming from the online Polidoxa community. The concept of "swarm intelligence" in nature has been primarily investigated regarding the achievement of goals like, for example, finding optimal paths to food and other tasks of relevant importance for a community of social insects. Recently this concept has inspired the exploration of new algorithms to solve optimization problems, for example the "Travelling Salesman Problem" (Kuo, Horng, Kao, Lin, Lee, Chen, Pan & Terano, 2010). As stated earlier, from the point of view of the communication issue, Polidoxa's main goal is the improvement of information quality, general knowledge and discussion. Although this cannot be formulated as an optimization problem (there is simply no ideal optimum), "swarm intelligence" can be still exploited as it will be presented in next sections.

At a macroscale level, is there any analogy between Polidoxa and a swarm intelligent system?

In Nature "swarms", as systems, have some common characteristics, which have been analysed by (Dorigo & Birattari, 2007):

- They are composed by many individuals
- These individuals are identical, or in some cases, they have some small variations
- The single individual has only local knowledge of the system

- The overall result of the system is the interaction between the independent agents or interacting with the environment in a stigmergic way
- The resulting system implements is self-organizing

In the next paragraphs the concept of stigmergy, which was introduced in this above list, will be analysed in more details. In Nature "swarm intelligence" can emerge from members with limited intellectual capabilities per se (such as communities of ants, termites, bees etc..), interacting in an in direct and asynchronous way. However, the potential performance of a community made by individuals having higher intelligence is still an open issue. The idea that an online community made of human groups can exhibit an intelligent behaviour has been investigated by some authors (Luo, Xia, Yoshida and Wang. 2009). It was concluded that these communities have their own characteristics, which are different from "Team Intelligence" and "Business Intelligence", normally the target of "intelligence modeling". This is because in a team, only a small number of participants is involved. Furthermore, in business there is normally a hierarchy, scarce flexibility and openness, which are instead characteristics of online communities, where there is more freedom and participation may not be constant. New participants can bring in new ideas and creativity to the online community they join, as well as interpretations, for example regarding facts. The process is of course bidirectional, because incoming knowledge is also outcoming knowledge, back to the members The obvious consequence of this bidirectional flow is that the "gatekeeping" process, is no longer necessary to choose the topics which are relevant for the mass. The system will self-organize in search of relevant issues and news. (Luo, Xia, Yoshida & Wang, 2009), have started to study what community intelligence is and when it can emerge, defining it as the result of a "triple interwoven network of knowledge network, human network and media network (technological network)"

Polidoxa platform, being potentially based on a large number of members, can be affected by the "swarm effect" and consequently show collective intelligence, due to the interaction, knowledge transfer and exchange between a massive number of participants belonging to different networks. Although this system can ideally perform like a neural network in a brain, the authors recognize that being communities focused on interests, opinions etc, the concept of a "Global brain" (Russel, 1993) may not suitable for online communities and the expression of "Supernetworks", networks of networks, may be more appropriate (Nagurney, Wakolbinger, 2005) .

As a community of users, Polidoxa will be based on a network of humans, a media network, the web and a knowledge network. Knowledge will be stored on the Web and it will be re delivered to the individuals, in a way which is very similar to the performance of stigmergic systems (Dorigo, Bonabeau & Theraulaz, 2000). It can then be concluded that Luo, Xia, Yoshida and Wang's model can indeed fit and describe how Polidoxa can perform and how a stigmergic behaviour can emerge from the collective intelligence of this online community.

**A Stigmergic Behavioural System as in Swarms**
As previously anticipated, the concept of "Stigmergy (Dorigo, Bonabeau and Theraulaz, 2000) has been regarded as significant in shaping the Polidoxa platform. The idea is inspired by Social Insect colonies, which are huge communities. To communicate they use a "face to face" communication system without the intervention of a centralized artificial medium (Miller, 2010). This system guarantees that information will never be centralized by a small colony subset. The knowledge sharing process works bottom-up, following the principles of democracy as described regarding the Agora. This principle seems to work efficiently in insects communities. Every time the information is passed on, the receiver checks who the sender is. If some information is ambiguous, the receiver stops the information flow and sends other insects to control which information is actually the correct one. As previously analysed, the human way to transfer and share information is different and it is normally influenced at best, or filtered, at worst, by mass media (TV, radio, newspaper, books, education system, etc…). Humans communicate in an unreliable way because they almost entirely rely upon mass media, bypassing every democratic principle and accepting a top down sharing of information. From a technical point of view, while a very strict group of people may have the potential to control mass information, this is impossible with the insects' communication model, which does not permit a centralized control. Consequently the authors believe that insect colonies behaviour may inspire

Polidoxa mechanism to work efficiently without any necessary centralized control, apart from technical assistance.

If Polidoxa aims at working like a stigmergic system, social interaction and networking are enhanced by the "collective intelligence", which will be superior to the sum of knowledge of individuals, as analysed in previous chapter. As a result of this, Polidoxa can consequently offer a platform for discussion which may contribute at elevating users' higher level of knowledge, criticism and consciousness (open source projects like Wikipedia, assuming there are no administrators, are examples of successful stigmergic systems). They work in a similar way to how social insect colonies build up a complex system to tell each other where to locate sources of food or picking up materials. Wikipedia is based on a collaborative system, without any external instruction, guidance or hierarchy. In the same way, Polidoxa users, as a colony of brains, can a) share information, b) interact with it, c) generate discussion, d) enhance the service itself, e) redefine how it will work in the future, etc. This happens as in a self-organizing system, which facilitates cooperative team work. This evolution from chaotic groups to self-organized users groups without any central guidance, will help in redefining of how information can be delivered, offering a real alternative to the traditional media top-down approach. The limitations imposed by the lack of users' guidance and hierarchy to meet the community goals, are possibly overcome by introducing a Holonic System functionality in Polidoxa.

## How Polidoxa as a Holonic System Addresses the Lack of Hierarchy

The concept of a holonic systems, firstly coined by Kostler (Koestler,1968) can be expressed in engineering terms as that of a system that is made up of autonomous units who are themselves (sub)systems, all acting in a cooperative way (Brennan, 2001). Although subject to the system's supra-hierarchy called holarchy, self-reliant units are characterised by a degree of independence, that aims at self-sustaining, stability and efficient use of resources (Calabrese, 2011). The intrinsic duality of a Holonic system, being simultaneously a "whole" and a "part" brings in a potentially new approach in how to implement the aims of the Polidoxa Platform.

An example of a holonic system and its duality is the human body; being it a whole system whose physical boundaries could be set as the skin that "senses" many, although not all the external "signals", it transfers these signals to the brain which interprets them and instructs the specialised sub-systems (such as organs, muscles, etc.) to perform the required action(s). The Holon, seen as a self-contained autonomous and cooperative entity, can be described also as a dissipative system in thermodynamics. A dissipative system (an Open System) is capable of exchanging energy and matter and interacts with the outer environment by means of its surroundings; any exchange of energy and/or matter results in the modification of its internal energy. These exchanges can be considered stimulus and they will produce a response that is managed by higher-level components "super-holons" and is transferred to lower-level components "sub-holons". A Holonic system is represented in Figure 5.

*Figure 5. A Holonic System composed by a Super-Holon, a lower level Holon and a further lower-level sub-holons. When stimulated by the outer environment the holon produces responses to sub-holons and then to the outer environment.*

If the stimulus that the Super-Holon detects from the external environment through suitable sensors can be of different nature, and not all may be beneficial, it is of primary interest to understand and implement how it is it possible to prevent malicious stimulus to affect the holon. This challenging task requires that stimulus, or Users from now on, are recognised as genuine ones after being quarantined before gaining access. Any recognised malicious user will then be expelled from the system and prevented from re-entry even if it changes "identity". The following paragraphs provide a methodological approach towards the holonic quarantine.

## Invited Users or Self-Candidacy Users

Polidoxa aims at redefining the Trust and Trust defines acceptance among users. Polidoxa is based on a holonic system which acts as a whole system and, simultaneously, as a cooperating set of sub-systems. Users' attempts to gain entry to the Polidoxa Platform need to be recognised not only as non-malicious, but they have to comply with the goals of a) being self-sustaining, b) increase system's stability and c) make efficient use of available resources (of the Polidoxa users' community). Quarantine is believed to be an effective way for any user to be accepted by the Polidoxa community; quarantine is performed by peers (other Polidoxa users) who are fully specialised and so capable of recognising a similar pair.

A very interesting example of such peer-reviewed activity is offered once again by nature. Studies on ants' communities demonstrated that once a colony member has been infected, its nest-mates perform a grooming activity toward the affected pair with the ultimate goal of guaranteeing the survival of the whole colony (Konrad, Vyleta, Theis, Stock, Tragust, Klatt, Drescher, Marr, Ugelvig, & Cremer, 2012). The tasks can be broken down in few steps and the ants behave in such a way that they 1) share part of the fungal infection in order lower its concentration from the severely affected individual to non-lethal values, 2) transfer (sharing) of low infection levels triggers the immune system of the grooming group and speeds up the healing process of the affected individual 3) the immunizing agent acts as a marker for the recognition of future occurrences of the infection. By doing so the community is preserved from identical future infections because the immunization information has become part of common knowledge (ibidem).

Social contact to pathogen-exposed individuals, or malicious users, enables immunisation of the entire colony, the Polidoxa Community, and this is applied to Polidoxa. A sequence of steps would be as follows: a) each user intervenes in quarantining a new user and is (or should be) capable of recognising any potential threat, b) acts to remove the malicious pathogen, c) keep memory of it and shares the information for social immunisation and d) the malicious user is marked and permanently banned from any future re-entry attempt.

## Trust as a Function of Network Activity

As shown previously, trust changes as a function of distance. Furthermore, trust changes over time as a function of network activity. The following parameters have been individuated as being relevant to update trust:

- For each user, the number of likes related to his posts: user popularity
- For each user, the ratio #Likes/#Dislike (with #A cardinality of set A) for that user
- Rate of activities (share, comments, like, dislike...) on a posted item within a temporal interval
- Number of private messages between the user and another user
- For each post of the user, the number of comments coming from another user
- Number of user comments to posts coming from another user
- Followers list
- Users that belong to subscribed groups
- Each group to which the user belong, number of the published posts on that group
- List of configurable keywords
- Favourites sites/blogs list
- Post labels
- Post frequencies
- RSS feed's list of the user and of all the first grade user's contacts –i.e., people directly connected with him – (configurable in case of extension to more than one level)

The following table synthetizes how trust has to be calculated at given time intervals.

*Table 1. How trust has to be calculated at given time intervals.*

All these parameters can be used to identify malicious users and non-trustworthy information in the style of Immune Network Systems but this goes beyond the scope of this work.

## POLIDOXA@TWITTER

Polidoxa@twitter (Chamot A., 2012) is the implementation of a simplified version of the Polidoxa's principles on top of the Twitter platform. Twitter (https://twitter.com) has many of the described characteristics of social networks like Facebook and LinkedIn. It has a simplicity which makes it a very good case study to experiment the ideas presented in this work, without worrying about unnecessary complications: information as it is presented in Facebook and LinkedIn is structured and varied (text, pictures, videos etc…); Twitter instead enables its users to send and read only text-based posts of up to 140 characters (the "tweets"). This makes easier to collect and store them for analysis purposes. Having Twitter mostly text information, only text-based analysis is necessary (no picture recognition or particular data analysis etc.…). Since the number of tweets exchanged in a given timeframe is much higher than the number of Facebook or LinkedIn posts, it is therefore faster to create a collection of relevant data (news). Furthermore, Tweets messages contain *hashtags* which are important to make the analysis more effective and efficient.

The remaining of this section will present an overview of the current prototype. Describing the implementation details is far out of the scope of this work and it would require a dedicated treatment. The prototype has been developed as a Twitter application that provides services to collect, store and analyze information gathered from Twitter. From a functional point of view, the architecture of the system can be divided in 3 main components: a *front-end application* acting as interface between the user and a *back-end application* that implements the data model and all the relative Polidoxa services. The *back-end application* interacts with a mysql *database* in which all the data is stored. The *front-end application* provides an access control model based on two different user roles: administrator and normal user. Let us focus on the functionalities offered to a normal user.

In order to use the Polidoxa application, each user must authenticate using his/her Twitter account, as shown in Figure 6.

*Figure 6. User authentication.*

If the authentication succeeds, then the user is redirected to a registration form that has to be filled in to complete the registration.

*Figure 7. User registration form.*

Finally, the application checks whether the twitter account has not been already registered. In case of success, a validation email is sent to the user with a link to validate the account. After validation, the registration is completed and the user can finally log-in.

## Static Trust

Each user's contact in a Polidoxa@twitter network is associated with a static parameter representing the trust value defined by the user for his/her relationship with that particular contact. Trust is a percentage with a default value of 50 %. The user can update this value at any time (Figure 8). Tweets are initially visualized/ranked/ordered according to this value.

*Figure 8. Static trust setup.*

## Dynamic Trust

Contacts that have a specific activities history have to be considered more relevant and have an automatic offset/boost of their trust values (for example a person with a default of 50% after some activity could rise to 55% and then 60 %). The dynamic trust is used to order result on the basis of the user network activities. This network activity information has been collected and stored in a back-end database. The dynamic trust is calculated for each contact using the formula shown in Figure 9.

*Figure 9. Trust formula.*

The formula contains the following parameters:

- Static Trust : User-defined value in the range 0..10
- Nbr_favorites: Number of tweets sent by the contact and favoured by the user. This number is multiply by a coefficient chosen by the administrator.
- Nbr_retweets: Number of tweets sent by the contact and retweeted by the user. This number is multiply by a coefficient chosen by the administrator.
- Nbr_mentions : Number of tweets sent by the user containing mentions referring this friend. This number is multiply by a coefficient chosen by the administrator.
- Nbr_FridayFollows : Number of tweets sent by the user containing FridayFollows referring this friend. This number is multiply by a coefficient chosen by the administrator
- Results_count : Number of tweets belonging to the friend matching the given search. This number is multiply by a coefficient chosen by the administrator.

The first four parameters are related to the activity occurring between the user performing the search and each of his/her contact. The idea is to give more importance to those contacts with whom the user interacted more. The last parameter is instead not specific to a contact but to a search: it corresponds to the number of matching items when a query is performed.

In order to decrease the dynamic trust value, the network activity of each user does not take into account items older than one full year. Therefore, if a user has not been interacting with a contact for some time, this will result in a decrement of the dynamic trust.

The coefficients appearing in the formula above can be set at any time only by the system administrator. This can be done throughout a dedicated interface, as shown in Figure 9.

*Figure 10. Configuration of coefficients for dynamic trust.*

The result of a search is displayed to the user as shown in Figure 10. The number of results is given at the top of the list of tweets. This list is composed of ordered tweets. For each of them the trust value is shown at the top right corner. Only the first 50 results are loaded. If the user scrolls down then the next 50 are added and son on, until no more results are available.

*Figure 11. Static trust: search sesults.*

Searches can be configured by means of the parameter search menu (Figure 11). Three basic funcionalities have been developed in the current version of the prototype: (1) one field allows to switch between static and dynamic trust (to order tweets); (2) a number of fields let the user specify the time range of the search; (3) finally, the last option gives the possibility to restrict the search to some specific friends only.

*Figure 12. Configuration of search parameters.*

## Static VS Dynamic

In this section a simple example is discussed to show how searches based on static and dynamic trust differ and how (and why) they actually generate different results. The example is based on database content which has been used to test the application. The content is not actually based on a prolonged and real use of the tool, but still it is realistic and it provides enough evidence to draw our conclusion.

Figure 10 shows the results of the search of the keyword "apple" under static trust. In particular, the static trust setup is the one depicted in Figure 8. As the screen-shot in Figure 10 shows, the results coming from the newspaper twitter account "ouest-france" are the first displayed. This is because the source is considered the most trustworthy by the user itself (55% of trust). The results are then ordered according to the static trust previously configured (Figure 8).

Let us assume now that the user decides to switch from static to dynamic trust in the search parameters and he/she runs the same search again. The result of the experiment is shown in Figure 12.

In this case the first results of the search is a tweet coming from "TechCrunch" and showing a trust value of 56.02%. This is due to the fact that the Twitter account associated with this user had more interactions with this account (accordingly to the formula showed above). Thus, despite the fact that the user might think that "ouest-france" is more trustworthy than "TechCrunch", his/her activities on the network show exactly the contrary. This ia a very interesting aspect showing how Polidoxa can provide informed recommendations based on social network analysis, in particular on interactions a users had with his network.

Overall and in the middle-run, this mechanism tend to correct users judgment (which sometime can be biased) by adapting the trust value according to accountable actions (retweets, mentions etc...) and not to personal judgment only. Analysis and simulation to show evidence of this specific fact are left as future works. For now, we would like to point out how the major impact of dynamic trust seems to be its potential to correct users' biased judgement by making use of recorded and undisputed interactions/activities data.

*Figure 13. Dynamic trust: search results.*

## RELATED WORKS

In this section we compare the Polidoxa idea with Google and Grouplens. PageRank is the parameter used by Google and it is based on the links received by a page and on the "authority" of certain pages. Thus, when a page is linked by another page with "authority", this gives more relevance to the page itself. The important question here is: how can we decide about the authority of a page? This is not clear and Google says nothing about it. Who works in SEO (Search Engine Optimization) — like one of the authors does — knows very well that "inlinks evaluation" (evaluation of links coming from other pages) is a process which lasts for months. This means that a page with qualitatively valuable information actually needs months to acquire some "authority". With Polidoxa, everything instead depends on the network's activity, without a delay of months but actually minutes. Relevance of information for Google is decided in a "black box" with a non-transparent process, and it can therefore be manipulated by SEO specialized agency (an online marketing branch which has the goal of bringing a page or document in search engines ranking top position). Polidoxa aims at offering a simple answer to this issue since relevance of information is determined by users' social network and it cannot be influenced by SEO agency. Polidoxa introduces a trust ranking algorithm where:

1. The user assigns a trust parameter (a numeric value/percentage) to each of his/her first level contact

2. The user's first level network determine a dynamic trust parameter on the basis of its activity (e.g., retweet, FF…)

Potentially (but not implemented here), users' indirect links (contacts of contacts) can also influence dynamic trust on the basis of their activity; in this case, how much the trust value is influenced would depend on the actual distance, according to some exponentially inverse law (as discussed in this work).

Users have now a unique instrument for searching information which values more thier direct connections, without limiting the use of traditional media or search engine. As a consequence, users are more willing to use their critical thinking when reading news and they are motivated to think about the sources and the process of news creation and dissemination. Indeed, all the filters created by the so-called "subject matter experts" of Grouplens (http://www.grouplens.org/biblio) are, in reality, not very transparent. For example, who decides who is an expert? Furthermore, an "expert" can be easily manipulated. With Polidoxa the "subject matter experts" is instead precisely decided by the users and not by an unknown, not better defined, external entity.
Polidoxa gives users the possibility of configuring their searches and the related ranking. It does not limit the general network activity but it gives the user a chance of monitoring the specific activity of his/her trusted network. The fundamental idea is that we tend to trust more the people we know and with these people we usually discuss more, get more feedback, interact more, etc. However, the possibility to follow famous people we do not directly know but, for some reason we trust, is not prevented. This is because a user may want to follow a distant person who is considered a role/spiritual model. Certainly, also in this virtual trusted network all the persuasion/influence mechanisms may still be valid and alter the trust relationship in a not obvious way. These aspects are described in detail by Cialdini (Cialdini, 2000).

Polidoxa users have the opportunity to be aware of the activity of the trusted network but still have to use their critical thinking to evaluate the information. This should give the opportunity to the "deep Web" (all that information not crawled by search engines) to eventually reach the Web surface. The Polidoxa ranking increases the quality of information, facilitates the discussion and could improve the lifestyle of participants simply exchanging information and sharing knowledge. Looking at the data of seo-scientist.com (http://www.seo-scientist.com) we discover that about 80% of the users just click the first three results given by a search engine. As a consequence, ranking of information is of extreme importance and offering a trust ranking based on the users activities is fundamental to offer qualitatively better results because that means improving the first three positions according to the user priorities and preferences. With Polidoxa the user and his/her trusted network influences the ranking and everybody has the chance to receive a customized and configurable ranking.

## FUTURE WORKS
A Polidoxa based news search engine is a promising idea to explore. The engine would be based on a configurable ranking algorithm. Users will be able to choose the sources from which the engine should retrieve the topics and ranking criteria may also be selected. The trustworthy social network allows following the information posted exclusively by trusted users on specific topics, which can be set. Other approaches can be found on the market, such as Google+, that support choices based on other people opinions. Polidoxa extends this idea by proposing a built-in search engine and by organizing people in a trustworthy social network where news positively evaluated by linked contacts, have a higher priority than the ones evaluated by indirect contacts. The higher the degree of separation, the lower the priority. The major difference with the Google algorithm is that Pagerank evaluates the link relationships of a document by looking at the entire Web, while Polidoxa evaluates the link relationships of the network community, giving more importance to the network activities within a shorter relational distance.

## CONCLUSION
This chapter contributes with several principles and technicalities to build a social platform to achieve collective intelligence via information sharing among trusted contacts. It also presents a Twitter-based implementation of a subset of these principles. Trust modeling, social networks, collective intelligence, algorithms and the relative motivations supported by literature in communication sciences are a quite inedited interdisciplinary blend, which has not been really investigates so far. We intend to pursue our investigation and move the human knowledge further on this topic

The fact that people tend to passively receive TV information without verifying it has been emphasized. The gatekeeping process of traditional media, although generally considered a safe and quality assuring medium, poses new risks when control over the information is becoming more and more centralized. Internet has an enormous potential to fix this issue, but the current instruments commonly used like Google and Facebook lack the most important concept in this field: they do not embed the notion of individual trustworthiness of a source. Polidoxa, instead, connects local knowledge making it accessible for everybody and it is conceived to promote public awareness and discussion in total freedom, like in an open piazza. Polidoxa is based on our philosophy: *"we believe first in what we can directly verify, then in what our closest contacts have verified. We doubt about what people we do not know say about things we have never seen (it does not matter if this is coming from official sources) until our network of trusted contacts allows us to trust it because it has been verified directly by them."* Today we tend not to verify mainstream information and this has the potential to become a growing issue in the future. Polidoxa may be an answer to this problem.

The same principles on which Polidoxa is based (collective intelligence, collaboration, stigmergy…) apply to the full implementation of the system too. We are indeed looking for potential collaborators interested in developing aspects of this project both at the theoretical/algorithmic level and at the software engineering/implementation level. We believe that only collective intelligence can create platforms for the exploitations of collective intelligence itself. We believe that cooperation, knowledge sharing between individuals and quality of life in general, can all be significantly improved by taking inspiration from nature.

## ACKNOWLEDGMENTS

The idea of Polidoxa slowly emerged among friends and its basic principles have been discussed and experimented over time with information sharing via other experimental platforms or, sometime, just emails. This topic has been discussed with several friends and colleagues before taking shape. All of them contributed substantially to the idea or the realization. We want to thank, in particular, Georgios Papageorgiou, Luca Chiarabini, Giuseppe Marraffa, Luca Ermini, Matteo Dall'Osso, Fabrizio Casalin and Chiara Succi. We also want to thank members of the Reconfiguration Interest Group and the Dependability Group at Newcastle University and the people involved in the EU FP7 DEPLOY Project (Industrial deployment of system engineering methods providing high dependability and productivity). A big acknowledgment to Antoine Chamot from DTU (for having collaborated in the implementation of the first prototype). Finally, colleagues and friends at UNU-IIST, Macau cannot be forgotten. This work has been partially supported by the SAFEHR project funded by Macao Science and Technology Development Fund.

## REFERENCES

Arendt, H. (1958), *Vita Activa. The Human Conditon*, University of Chicago Press.

Asch, S. (1955). *Opinions and social pressure,* Scientific American, Vol. 193 (1955), pp. 31-35

Auletta, K. (2010), *Googled: The End of the World as We Know It*, Virgin Book, London.

Beach,A., Gartrell & M., Han, R. (2009), *Solutions to security and privacy issues in mobile social networking.* In CSE (4), pages 1036–1042.

Beni G. & Wang J. (1989), *Swarm Intelligence in Cellular Robotic System*s, Proceed. NATO Advanced Workshop on Robots and Biological Systems, Tuscany, Italy

Bonneau J., Anderson J. & Church L. (2009), *Privacy suites: shared privacy for social networks,* SOUPS 2009, Proceedings of the 5th Symposium on Usable Privacy and Security

Bonabeau E., Dorigo M. & Theraulaz G. (1999), *Swarm Intelligence: from Natural to Artificial Systems*. Oxford University Press, OUP USA Inc., New York

Brennan R.W. (2001), *Holonic and multi-agent systems in industry*. The Knowledge Engineering Review ", 16(04): 375-381.

Calabrese, M. (2011), Hierarchical-Granularity Holonic Modelling. PhD Thesis, University of Milan, Italy

Carr, N. (2010), *The Shallows: What the Internet Is Doing to Our Brains,* W. W. Norton & Co.

Christakis, N.A. & Fowler, J.H. (2009), Connected: The Surprising Power of Our Social Network and How They Shape Our Lives. New York : Little, Brown and Co.

Chun, W. H. K. & Keenan, T. W. (2005)*, New Media, Old Media: A History and Theory Reader*, Routledge, New Ed edition.

Cutillo L.A., Molva R. & Önen M. (2011), *Safebook: A distributed privacy preserving online social network*. In WOWMOM, pages 1–3.

Cialdini R. (2000), Influence*: Science and practice,* Allyn & Bacon, Boston, Massachusetts.

Ding J., Cruz I. & Li C. (2009), *A formal model for building a social network*. In SOLI. IEEEXplore

Dorigo M., Bonabeau E. & Theraulaz G. (2000), A*nt algorithms and stigmergy*, Future Generation Computer Systems, v.16 n.9, p.851-871.

Dorigo M. & Birattari M. (2007), *Swarm intelligence,* Scholarpedia, 2(9):1462.

Entman, R.M. (1989), *How the media affect what people think: an information processing approach*. The Journal of Politics, 51 (2), 347-370

Fong P. W. L., Anwar M. M. & Zhao Z. (2009), *A privacy preservation model for facebook-style social network systems*. In ESORICS, pages 303–320.

Joslyn C., Rocha L., Smith S., Johnson N.L., Rasmussen S. & Kantor M. (1998), *Symbiotic intelligence: Self-organizing knowledge on distributed networks driven by human interaction*, In C. Adami, et al. (Eds.), *Artificial Life VI*. Cambridge, Mass., MIT Press.

Koestler A. (1968), *The ghost in the machine. New York,* Macmillan.

Konrad, M., Vyleta, M. L., Theis, F. J., Stock, M., Tragust, S., Klatt, M., Drescher, V., Marr, C., Ugelvig, L. V. & Cremer, S. (2012), *Social Transfer of Pathogenic Fungus Promotes Active Immunisation in Ant Colonies*. PLoS Biol 10(4): e1001300, 2012.

Kuo I., Horng S., Kao T., Lin T., Lee C., Chen Y., Pan Y.I. & Terano T. (2010), *A hybrid swarm intelligence algorithm for the travelling salesman problem,* Expert Systems, volume 27, number 3, 2010, pages 166-179.

Luo, S., Xia, H., Yoshida, T., Wang, Z., 2009, *Toward collective intelligence of online communities: A primitive conceptual model,* Journal of Systems Science and Systems Engineering, vol. 18, pp. 203-221, 2009.

Maheswaran M., Cheong Tang H. & Ghunaim A. (2007), *Towards a gravity-based trust model for social networking systems*, Distributed Computing Systems Workshops, International Conference on, 0:24, 2007.

Maurer H. & Kolbitsch J. (2006), *The transformation of the web: How emerging communities shape the information we consume*, in "Journal of Universal Computing Science", 12(2):187–213.

Mazzara M., Marraffa A., Biselli L., & Chiarabini L. (2011), *The Polidoxa Shift: a New Approach to Social Networks*, in "Journal of Internet Services and Information Security" (JISIS), 1(4):74-88.

Mazzara M., Marraffa A., Biselli L., & Chiarabini L., (2011)*, Polidoxa: a Synergic Approach of a Social Network and a Search Engine to Offer Trustworthy News*, in "International Workshop on TRUstworthy Service-Oriented Computing (INTRUSO 2011), Copenhagen, Denmark.

Mayer R.C., Davis J. H & Schoorman F.D, (1995), *An integrative model of organizational trust.* Academy of Management Review. 20 (3), 709-734.

Mccombs M. & Shaw D. (1972), *The agenda-setting function of mass media*, in "Public Opinion Quarterly", 36:176–187, University of Chicago Press, Chicago

Mccombs M. (2004), *Setting the Agenda: the mass media and public opinion*, Blackwell Publishing, New York.

Miller P. (2010), *Smart Swarm: Using Animal Behaviour to Organise Our World,* Collins

Nagurney & T. Wakolbinger (2005), *Supernetworks: An introduction to the concept and its applications with a specific focus on knowledge supernetworks*. International Journal of Knowledge, Culture and Change Management.

Noelle-Neumann E. (1974), *The spiral of silence: a theory of public opinion*, in " Journal of Communication", 24:43–51, USA.

Parikka, J. (2010), *Insect Media. An Archaeology of Animals and Technology*, University of Minnesota Press.

Rosalind, T. (1992), *Literacy and Orality in Ancient Greece*, Cambridge University Press, Cambridge.

Russell P. (1983), *The Global Brain: Speculations on the Evolutionary Leap to Planetary Consciousness*. Houghton Mifflin, Boston, MA

Sampson, T. D. (2012), *Virality. Contagion Theory in the Age of Networks,* University of Minnesota Press.

Shuangling L., Haoxiang X., Taketoshi Y. & Zhongtuo W. (2009), *Toward collective intelligence of online communities: a primitive conceptual model*. Journal of Systems Science and Systems Engineering.

Steven J., (2002), *Emergence: The Connected Lives of Ants, Brains, Cities and Software*, Penguin.

Travers J., Milgram S. (1969), *An Experimental Study of the Small World Problem,* Sociometry, 1969, volume 32, pages 425—443

White, D. M. (1950), *The "gate-keeper": A case study in the selection of news*, Journalism Quarterly, 27, 383-390.

## ADDITIONAL READINGS

Bruno N. (2010), *Tweet first, verify later? How real-time information is changing the coverage of worldwide crisis events'*, Reuters Institute for the study on journalism
http://reutersinstitute.politics.ox.ac.uk/fileadmin/documents/Publications/fellows__papers/2010-2011/TWEET_FIRST_VERIFY_LATER.pdf

Clayman S. & Reisner A.(1998), *Gatekeeping in Action: Editorial Conferences and Assessments of Newsworthiness*, American Sociological Review 63: 178–99.

Fidler R. (1997), *Mediamorphosis: Understanding New Media.* Thousand Oaks, CA: Pine Forge.

Griffith B. (2003), *High-Tech Multitasking: Fan Chat is Message to Media'*, The Boston Globe 21, December: C10.

Haesen R., Baesens B., Martens D., De Backer M. & Holvoet T. (2006), *Ants constructing rule-based classifiers,* in *"Technical report"*, Katholieke Universiteit Press, Leuven.

Katz, E., Lazarsfeld, P.F. (1956), *Personal Influence: the Part Played by People in the Flow of Mass Communication's,* Columbia Press, Glencoe.

Lewis, J., (1992), *A History of Ancient Greece.The Glory That Was Greece. The Agora,* International World History Project, at http://history-world.org/agora.htm, (accessed August 23, 2012).

Livingston S. & Lance Bennett W. (2003), *Gatekeeping, Indexing, and Live-Event News: Is Technology Altering the Construction of News?* Political Communication 20: 363–80

McQuail, D., (2010), *Mass Communication Theory* (sixth edition, first edition: 1987), Sage, London.

Morozov, E. (2011), *The Net Delusion: The Dark Side of Internet Freedom*, PublicAffairs.

Walsh P. (2003), *The Withered Paradigm: The Web, the Expert and the Information Hegemony*
In Henry Jenkins and David Thornurn (eds) Democracy and New Media, pp. 365–72. Cambridge, MA: MIT Press

Chamot A. (2012), *Prototype implementation of a social network application for the Polidoxa Project*, Master's thesis, Technical University of Denmark, 2012

## KEY TERMS AND DEFINITIONS

**SOCIAL NETWORK**: A service (generally web-based) allowing individuals to construct a (semi) public profile and to define a set of other users with whom they share a privileged first-level connection. Among first-level connections messaging and other forms of interactions are possible.

**COLLECTIVE INTELLIGENCE**: A shared or group intelligence emerging from the collaboration and/or competition of many individuals.

**AGORA**: The central spot in ancient Greek city-states. The literal meaning of the word is "gathering place" or "assembly". The agora was the centre of athletic, artistic, spiritual and political life of the city.

**GATEKEEPING**: The process through which information is filtered for dissemination, whether for publication, broadcasting, the Internet, or some other mode of communication.

**TRUST**: A measurement/metric of the degree to which one user trusts the activities of another.

**STIGMERGY**: A mechanism of indirect coordination between agents or actions. The principle is that the trace left in the environment by an action stimulates the performance of a next action, by the same or a different agent. In that way, subsequent actions tend to reinforce and build on each other, leading to the spontaneous emergence of coherent, apparently systematic activity.

**HOLONIC SYSTEM**: A systems composed of autonomous units who are themselves (sub) systems, all acting in a cooperative way.